\documentclass[a4paper,12pt,abstracton,bibtotoc,
headsepline,footexclude,headexclude
]{scrartcl}
\usepackage{graphicx}
\usepackage{wrapfig}
\usepackage{natbib}
\usepackage{hyperref}

\bibpunct{[}{]}{,}{n}{}{,}

\newcommand{\theTitle}{Tagging, Folksonomy \& Co -- Renaissance of Manual Indexing?}
\title{\theTitle\footnote{
Submitted to the 10th International Symposium for Information Science, Cologne.}}
\author{\textit{Jakob Vo\ss}}
\date{\normalsize January 2007}
\publishers{\normalsize
Common Library Network GBV\\
G\"ottingen, Germany\\
jakob.voss@gbv.de
}

\begin{document}
\maketitle

\begin{abstract}
\noindent This paper gives an overview of current trends in manual indexing on 
the Web. Along with a general rise of user generated content there are more 
and more tagging systems that allow users to annotate digital resources with 
tags (keywords) and share their annotations with other users. Tagging is 
frequently seen in contrast to traditional knowledge organization systems or 
as something completely new. This paper shows that tagging should better be 
seen as a popular form of manual indexing on the Web. Difference between 
controlled and free indexing blurs with sufficient feedback mechanisms.
A revised typology of tagging systems is presented that includes different 
user roles and knowledge organization systems with hierarchical relationships 
and vocabulary control.  A detailed bibliography of current research in
collaborative tagging is included.
\end{abstract}

{\small {\sfb Free Keywords:} 
tagging, indexing, knowledge organization, typology}

{\small {\sfb ACM Computing Classification:}
{\tt H.3.1.} Content Analysis and Indexing}

{\small {\sfb arXiv/CoRR Subject Classification:}
{\tt IR.} Information Retrieval, 
{\tt DL.} Digital Libraries}

{\small {\sfb JITA Classification:}
{\tt IC.} Index languages, processes and schemes}

\pagebreak

\section{Introduction}
The World Wide Web, a framework originally designed for information 
management \cite{BernersLee1989}, has long ago become a heterogeneous, 
exponentially growing mass of connected, digital resources. After first, 
unsuccessful attempts to classify the Web with traditional, intellectual 
methods of library and information science, 
the standard to search the Web is now fulltext indexing -- most of all made 
popular by Google's PageRank algorithm. The success of such automatic 
techniques is a reason why ``many now working in information retrieval 
seem completely unaware that procedures other than fully automatic ones have 
been applied, with some success, to information retrieval for more then 
100 years, and that there exist an information retrieval literature beyond 
that of the computer science community.''\cite{Lancaster2003}
However in the recent years there is a renaissance of manual subject indexing 
and analysis: Structured metadata is published with techniques like RSS, 
OAI-PMH, and RDF. OpenSearch\footnote{To gain an insight on RSS, OAI-PMH, RDF, 
and OpenSearch see \url{http://en.wikipedia.org/}.}
and browser search plugins allow it to aggregate specialised search engines.
Last but not least many popular social software systems contain methods to 
annotate resources with keywords. This type of manual indexing is called 
\textit{tagging} with index terms referred to as \textit{tags}. Based on 
\cite{Marlow2006} this paper presents a revised typology of tagging 
systems that also includes systems with controlled and structured 
vocabularies. Section~\ref{TaggingIntroduction} gives a short introduction 
to current tagging systems and its research. Afterwards 
(section~\ref{IndexingProcess}) theory of subject indexing is 
pictured with the indexing process, typology of knowledge organization 
systems, and an unconventional look at vocabulary control. In 
section~ \ref{TaggingTypology} the typology of tagging systems is 
presented with conclusion in section~\ref{Conclusion}.

\section{Tagging systems on the rise}\label{TaggingIntroduction}
Tagging is referred to with several names: \textit{collaborative tagging}, 
\textit{social classification}, \textit{social indexing}, \textit{folksonomy} 
etc. The basic principle is that end users do subject indexing instead of 
experts only, and the assigned tags are being shown immediately on the Web. 
The number of websites that support tagging has rapidly increased since 2004. 
Popular examples are del.icio.us (\url{http://del.icio.us}), furl 
(\url{http://furl.net}), reddit (\url{http://reddit.com}), and 
Digg (\url{http://digg.net}) for bookmarks \cite{Hammond2005} and 
flickr (\url{http://flickr.com}, \cite{Winget2006}) for photos.
Weblog authors usually tag 
their articles and specialized search engines like Technorati 
(\url{http://technorati.com/}) and RawSugar (\url{http://rawsugar.com}) 
make use of it. But tagging is not limited to simple keywords only: 
BibSonomy (\url{http://bibsonomy.org}, 
\cite{Hotho2006Emergent, Hotho2006Entstehen}), 
Connotea (\url{http://connotea.org}, \cite{Lund2005}), 
CiteULike (\url{http://citeulike.org/}), 
and LibraryThing (\url{http://librarything.com})  
allow users to manage and share bibliographic metadata on the Web 
(also known as \textit{social reference managing} or 
\textit{collaborative cataloging}).
Additionally to librarian's subject indexing the University of Pennsylvania 
Library allows users to tag records in their online catalog since 2005 
(\url{http://tags.library.upenn.edu/}). Other systems to tag bibliographic
data are LibraryThing (\url{http://www.librarything.com}) and Amazon's tagging
feature (\url{http://amazon.com/gp/tagging/cloud/}). The popular free 
encyclopedia Wikipedia contains so called categories that 
are used as hierarchical tags to order the articles by topic \cite{Voss2006}.
Apart from social software there is also a rise of manual indexing in 
other fields \cite{Wright2005, Maislin2005}.

The details of tagging vary a lot but all applications are designed to 
be used as easy and as open as possible. Sometimes the greenness in 
theory of users and developers let you stumble upon known problems like 
homonyms an synonyms but on the other hand unloaded trial and error 
has led to many unconventional and innovative solutions.

\subsection{Research on Tagging}

The astonishing popularity of tagging led some even claim that it would 
overcome classification systems \cite{Shirky2005}, but it is more likely 
that there is no dichotomy \cite{Crawford2006} and tagging just stresses 
certain aspects of subject indexing. Meanwhile serious research 
about collaborative tagging is growing --- hopefully it will not have 
to redo all the works that has been done in the 20th century. At the 
15th World Wide Web Conference there was a \textit{Collaborative Web Tagging
Workshop}\footnote{\url{
http://www.rawsugar.com/www2006/taggingworkshopschedule.html}}.
The 17th SIG/CR Classification Research Workshop was about 
\textit{Social Classification}\footnote{\url{http://www.slais.ubc.ca/users/sigcr/events.html}}.
One of the first papers on folksonomies is \cite{Mathes2004}.
Shirky's paper \cite{Shirky2005} has reached huge impact. It is probably 
outdated but still worth to read. A good overview until the 
beginning of 2006 is given in \cite{Macgregor2006}. Some papers that deal 
with specific tagging systems are cited at the beginning of this section 
at page~\pageref{TaggingIntroduction}. Trant and Smith describe the 
application of tagging in a museum \cite{Trant2006a, Trant2006b, Smith2006}. 
Other works focus on 
tagging in enterprises \cite{Farrell2006, John2006, Damianos2006, Millen2006} 
or knowledge management \cite{Wu2006}. Another application is tagging 
people to find experts \cite{Bogers2006, Farrell2006}.
Mathematical models of tagging are elaborated in \cite{Tosic2006, Lambiotte2005}.
The usual model of tagging is a tripartite graph, the nodes being resources, 
users, and tags \cite{Lambiotte2005}. 
Several papers provide statistical analysis of tagging over time and evolution 
of tagging systems \cite{Kowatsch2007, Cattuto2006, Dubinko2006, Hotho2006Trend, Lin2006}.
Tagging behaviour is also topic of Kipp and Campbell~\cite{Kipp2006} 
and Feinberg~\cite{Feinberg2006}.
Types of structured and compound tagging are analyzed in 
\cite{BarIlan2006, Tonkin2006, Guy2006}.
Like in traditional scientometrics you can find communities and 
trends based on tagging data \cite{Jaeschke2006, Hotho2006Trend}.
Voss~\cite{Voss2006} finds typical distributions among different 
types of tagging systems and compares tagging systems with 
traditional classification and thesaurus structures.
Tennis~\cite{Tennis2006} uses framework analysis to compare 
social tagging and subject cataloguing.
Tagging is manual indexing instead of automatic indexing.
Ironically a focus of research is again on automatic systems 
that do data mining in tagging data
\cite{Aurnhammer2006, Hotho2006Emergent, Hotho2006Information, Schmitz2006Ontology, Schmitz2006Mining}.
Heymann and Garcia-Molina\cite{Heymann2006} presented an algorithm to
automatically generate hierarchies of tags out of flat, aggregated 
tagging systems with del.icio.us data. Similar approaches are used 
by Begelman et al.\ \cite{Begelman2006} and Mika et al.\ \cite{Mika2005}.
Research on tagging mostly comes from computer science and 
library science --- obviously there is a lack of input from 
psychology, sociology, and cognitive science in general 
(an exception from philosophy is Campell~\cite{Campbell2006} 
who applies Husserl's theory of phenomenology to tagging).


\pagebreak
\section{The indexing process}\label{IndexingProcess}
\begin{wrapfigure}{r}{8cm}
\begin{center}
\includegraphics{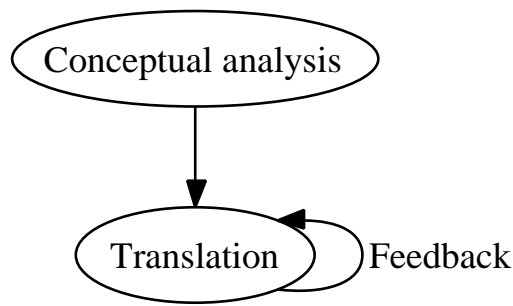}
\end{center}
\caption{Steps in subject indexing}
\label{indexingProcess}
\end{wrapfigure}
The main purpose of subject indexing is to construct a representation
of a resource that is being tagged. According to Lancaster
\citep[chapter 2]{Lancaster2003}
subject indexing involves two steps: conceptual analysis and translation
(see figure \ref{indexingProcess}). These are intellectually separate although
they are not always clearly distinguished. The semiotic triangle 
can be applied to indexing to demonstrate the distinction between object 
(resource), concept (what the resource is about), and symbol (set of tags to 
represent the resource). \textit{Conceptual analysis} involves deciding on 
what a resource is about and what is relevant in particular.
Note that the result of conceptual analysis heavily depends on the needs
and interests of users that a resource is tagged for --- different people
can be interested in different aspects.
\textit{Translation} is the process
of finding an appropriate set of index terms (tags) that represent the
substance of conceptual analysis. Tags can be extracted from the resource 
or assigned by an indexer. Many studies have shown that high consistency 
among different indexers is very difficult to achieve and affected by many 
factors \citep[chapter 3]{Lancaster2003}. One factor is control 
of the vocabulary that is used for tagging. Synonyms (multiple words and 
spellings for the same concept) and homonyms/homographs (words with different 
meanings) are frequent problems in the process of translation. A controlled 
vocabulary tries to eliminate them by providing a list of preferred and 
non-preferred terms, often together with definitions and a semantic structure. 
Controlled vocabularies are subsumed as knowledge organization systems (KOS) 
\cite{Zeng2004}. These systems have been studied and developed in library and 
information science for more then 100 years. Popular examples are the Dewey 
Decimal Classification, Ranganathan's faceted classification, and the first 
thesauri in the 1960s. Beginning with the 1950s library and information 
science has lost its leading role in the development of information retrieval 
systems and a rich variety of KOS has come into existence. However it is one of 
the constant activities of this profession to summarize and evaluate the 
complexity of attempts to organize the world's knowledge.

\subsection{Typology of knowledge organization systems}
Hodge, Zeng, and Tudhope \cite{Hodge2000, Zeng2000, Tudhope2006} distinguish 
by growing degree of language control and growing strength of semantic 
structure: term list, classifications and categories, and relationship groups.
Term lists like authority files, glossaries, gazetteers, and dictionaries 
emphasize lists of terms often with definitions. Classifications and 
categories like subject headings and classification schemes (also known 
as taxonomies) emphasize the creation of subject sets. Relationships 
groups like thesauri, semantic networks, and ontologies emphasize 
the connections between concepts. Apart from the training of what now 
may be called \textit{ontology engineers} the theoretical research on 
knowledge organization systems has had little impact on technical 
development. Only now common formats are being standardized with 
SKOS\footnote{\url{http://www.w3.org/2004/02/skos/}}, the microformats 
movement\footnote{\url{http://microformats.org/}} and other initiatives.
Common formats are a necessary but not sufficient condition for 
interoperability among knowledge organization systems --- an important 
but also frequently underestimated task \cite{Zeng2004, Mayr2006}.




\subsection{Vocabulary control and feedback}

In the process of indexing the controlled vocabulary is used to supply 
translation via \textit{feedback} (figure \ref{indexingProcess}). The 
indexer searches for index terms supported by the structure of the knowledge
organization systems until he finds the best matching tag. Also search 
is supported by the structure of the knowledge organization systems. 
Collaborative tagging also provide feedback. 

A special kind of tagging system is the category system of Wikipedia. 
The free encyclopedia is probably the first application of collaborative 
tagging with a thesaurus \cite{Voss2006}. The extend of contribution in 
Wikipedia is distributed very inhomogeneously (more precise it is a power 
law \cite{Voss2005}) --- this also applies for the category system. 
Everyone is allowed to change and add categories but most authors only 
edit the article text instead of tagging articles and even less authors 
change and add the category system. Furthermore each article is not 
tagged independently by every user but users have to agree on a single 
set of categories per article. So tagging in Wikipedia is somewhere between 
indexing with a controlled vocabulary and free keywords. Most of the time 
authors just use the categories that exist but they can also switch to 
editing the vocabulary at any time. The emerging system may look partly
chaotic but rather useful. With a comparison of Wikipedia and the 
AGROVOC\footnote{\url{http://www.fao.org/agrovoc}} thesaurus 
Milne et al.\ \cite{Milne2006} show that domain-specific thesauri 
can be enriched and created with Wikipedia's category and link structure.

We can deduce that the border between free keyword tagging on the one 
hand and tagging with a controlled vocabulary is less strict. Although 
most tagging systems do not implement vocabulary control there is almost 
always some feedback that influences tagging behaviour towards consensus: 
the Folksonomy emerges \cite{Mathes2004}. The phenomena is also known 
as \textit{emergent semantics} or \textit{Wisdom of the crowds} 
(But you should keep in mind that masses do not always act wise --
see Lanier's critic of `Digital Maoism' \cite{Lanier2006}).

\pagebreak

\section{Typology of tagging systems}\label{TaggingTypology}
Based on Marlow's taxonomy of tagging systems\cite{Marlow2006} I 
provide a revised typology. The following key dimensions do not 
represent simple continuums but basic properties that should be 
clarified for a given tagging system --- so they are presented 
here as questions.

\begin{description}

\item[Tagging Rights] Who is allowed to tag resources? Can any user
tag any resource or are there restrictions? Are restrictions based
on resources, tags, or users? Who decides on restrictions? Is there
a distinction between tags by different types of users and resources?

\item[Source of Resources] Do users contribute resources and have
resources been created or just supplied by users? Or do users tag
resources that are already in the system? Who decides which resources 
are tagged?

\item[Resource Representation] What kind of resource is being tagged?
How are resources presented while tagging (autopsi principle)?

\item[Tagging Feedback] How does the interface support tag entry? 
Do users see other tags assigned to the resource by other users or 
other resources tagged with the same tags? Does the system suggest 
tags and if so based on which algorithms? Does the system reject 
inappropriate tags?

\item[Tag Aggregation] Can a tag be assigned only once to a resource 
(set-model) or can the same tag can be assigned multiple times 
(bag-model with aggregation)?

\item[Vocabulary control]: Is there a restriction on which tags to use
and which tags not to use? Are tags created while tagging or is management
of the vocabulary a separated task? Who manages the vocabulary, how
frequently is it updated, and how are changes recorded?

\item[Vocabulary Connectivity] Are tags connected with relations? Are
relations associative (authority file), monohierarchical (classification
or taxonomy), multihierarchical (thesaurus), or typed (ontology)? Where
do the relations come from? Are relations limited to the common vocabulary
(precoordination) or can they dynamically be used in tagging 
(postcoordination with syntactic indexing)?

\item[Resource Connectivity] How are resources connected to each other 
with links or grouped hierarchically? Can resources be tagged on 
different hierarchy levels? How are connections created?

\item[Automatic Tagging] Is tagging enriched with automatically created tags 
and relations (for instance file types, automatic expansion of terms etc.)?

\end{description}

The analysis shows that the classic tripartite model of tagging with 
resources, users, and tags is too simplified to cover the variety of 
tagging system. Depending on the application you can distinguish 
different kind of resources, tags, and users. At least you 
should distinguish four user roles:

\begin{enumerate}
\item {\sfb Resource Author} A person that creates or edits a resource
\item {\sfb Resource Collector} A person that adds a resource to a tagging
system
\item {\sfb Indexer or Tagger} A person that tags resources
\item {\sfb Searcher} A person that uses tags to search for resources
\end{enumerate}

In most systems some of the roles overlap and people can fullfill different 
roles at different times (large libraries may be a counterexample). For 
instance the author of a private blog combines 1, 2, and 3,
a user of del.icio.us combines either 2 and 3 (tagging a new webpage) 
or 3 and 4 (copying a webpage that someone else has already tagged), 
and a  Wikipedia author combines either 1 and 2 (new articles) or 
1 and 3 (existing articles).

\section{Conclusion}\label{Conclusion}
The popularity of collaborative tagging on the Web has resurged 
interest in manual indexing. Tagging systems encourage users to 
manually annotate digital objects with free keywords and share 
their annotations. Tags are directly assigned by anyone who likes 
to participate. The instant visibility is motivation and helps to 
install feedback mechanism. Through feedback the drawbacks of 
uncontrolled indexing are less dramatic then in previous systems 
and the border between controlled and free indexing starts 
to blur. Vocabulary control and relationships between index terms 
should not be distinctive features of tagging systems and traditional 
knowledge organization systems but possible properties of manual 
indexing systems. Further research is needed to find out under which 
circumstances which features (for instance vocabulary control) are 
needed and how they influence tagging behaviour and evolution of the
tagging system. The typology of tagging systems that was presented 
in section \ref{TaggingTypology} combines all of them. The possibility 
to allow non-experts to add keywords has made collaborative 
tagging so popular --- but it is nothing fundamentally new.
Perhaps the most important feature of tagging systems on the Web 
is its implementation or how Joseph Busch entitled his keynote 
speech at the ASIST SIG-CR workshop: ``It's the interface, stupid!''
Today's tagging websites make many traditional knowledge organization 
systems look like stone age technique: effective but just too
uncomfortable. Some of the costly created thesauri and classifications 
are not even accessible in digital form at all (because of licensing
issues grounded in a pre-digital understanding of copyright or because 
of a lack of technological skills)! But also computer scientists 
tend to forget that a clever interface to support tagging can be 
worth much more than any elaborated algorithm. Anyway the art of 
creating interfaces for developed tagging systems is still in its 
infancy. Knowledge organization will always need manual input so it 
is costly to manage --- but Wikipedia showed that groups of volunteers 
can create large knowledge resources if a common goal and the right 
toolkit exist! And obviously there is not one way of indexing that fits 
for all applications. Collaborative Tagging is neither the successor 
of traditional indexing nor a short-dated trend but 
--- like Tennis \cite{Tennis2006} concludes --- a catalyst for 
improvement and innovation in indexing.

\setlength{\bibsep}{0.2em plus 0.3em} 
\bibliographystyle{plainurl}
\bibliography{isi2007voss}

\end{document}